\documentclass[twocolumn]{aastex63}
\usepackage{comment}
\usepackage{lineno}
\usepackage{grffile}
\usepackage{amsmath} 
\usepackage{color}
\usepackage{hyperref}
\usepackage{nameref}
\usepackage{enumitem} % added on August 9

\newcommand\name{COS-2987030247}
\newcommand\shortname{COS-2987}
\newcommand{\oiii}{[O\ {\sc iii}]}
\newcommand{\oii}{[O\ {\sc ii}]}
\newcommand{\cii}{[C\ {\sc ii}]}
\newcommand{\ciii}{C\ {\sc iii}]}
\newcommand{\cliii}{[Cl\ {\sc iii}]}
\newcommand{\nii}{[N\ {\sc ii}]}
\newcommand{\niv}{N\ {\sc iv}]}
\newcommand{\sii}{[S\ {\sc ii}]}
\newcommand{\ha}{H$\alpha$}
\newcommand{\hb}{H$\beta$}
\newcommand{\lsun}{$L_{\rm \odot}$}

\newcommand{\temp}{$T_{\rm e}$}
\newcommand{\dens}{$n_{\rm e}$}
\newcommand{\av}{$A_{V}$}

\submitjournal{ApJL}
\received{2025/July/3}
\accepted{??}

\shorttitle{ISM Physical Properties of \name}
\shortauthors{Usui et al.}
\graphicspath{{./}{figures/}}

\begin{document}
\title{\large RIOJA. JWST and ALMA unveil the inhomogeneous and complex ISM structure in a star-forming galaxy at $z=6.81$}
\correspondingauthor{Takuya Hashimoto}
\email{hashimoto.takuya.ga@u.tsukuba.ac.jp}
\author{Mitsutaka Usui}
\affiliation{Division of Physics, Faculty of Pure and Applied Sciences, University of Tsukuba, Tsukuba, Ibaraki 305-8571, Japan}
\author[0000-0003-4985-0201]{Ken Mawatari}
\affil{Waseda Research Institute for Science and Engineering, Faculty of Science and Engineering, Waseda University, 3-4-1 Okubo, Shinjuku, Tokyo 169-8555, Japan}
\affiliation{Department of Pure and Applied Physics, School of Advanced Science and Engineering, Faculty of Science and Engineering, Waseda University, 3-4-1 Okubo, Shinjuku, Tokyo 169-8555, Japan}
\affiliation{Division of Physics, Faculty of Pure and Applied Sciences, University of Tsukuba, Tsukuba, Ibaraki 305-8571, Japan}
\affiliation{Tomonaga Center for the History of the Universe (TCHoU), Faculty of Pure and Applied Sciences, University of Tsukuba, Tsukuba, Ibaraki 305-8571, Japan}
\author[0000-0002-7093-1877]{Javier \'Alvarez-M\'arquez}
\affiliation{Centro de Astrobiolog\'{\i}a (CAB), CSIC-INTA, Ctra. de Ajalvir km 4, Torrej\'on de Ardoz, E-28850, Madrid, Spain}
\author[0000-0002-0898-4038]{Takuya Hashimoto}
\affiliation{Division of Physics, Faculty of Pure and Applied Sciences, University of Tsukuba, Tsukuba, Ibaraki 305-8571, Japan}
\affiliation{Tomonaga Center for the History of the Universe (TCHoU), Faculty of Pure and Applied Sciences, University of Tsukuba, Tsukuba, Ibaraki 305-8571, Japan}
\author[0000-0001-6958-7856]{Yuma Sugahara}
\affiliation{Waseda Research Institute for Science and Engineering, Faculty of Science and Engineering, Waseda University, 3-4-1 Okubo, Shinjuku, Tokyo 169-8555, Japan}
\affiliation{Department of Pure and Applied Physics, School of Advanced Science and Engineering, Faculty of Science and Engineering, Waseda University, 3-4-1 Okubo, Shinjuku, Tokyo 169-8555, Japan}
\author[0000-0001-8442-1846]{Rui Marques-Chaves}
\affiliation{Geneva Observatory, Department of Astronomy, University of Geneva, Chemin Pegasi 51, CH-1290 Versoix, Switzerland}
\author[0000-0002-7779-8677]{Akio K. Inoue}
\affiliation{Waseda Research Institute for Science and Engineering, Faculty of Science and Engineering, Waseda University, 3-4-1 Okubo, Shinjuku, Tokyo 169-8555, Japan}
\affiliation{Department of Pure and Applied Physics, School of Advanced Science and Engineering, Faculty of Science and Engineering, Waseda University, 3-4-1 Okubo, Shinjuku, Tokyo 169-8555, Japan}
\author[0000-0002-9090-4227]{Luis Colina}
\affiliation{Centro de Astrobiolog\'{\i}a (CAB), CSIC-INTA, Ctra. de Ajalvir km 4, Torrej\'on de Ardoz, E-28850, Madrid, Spain}
\author[0000-0001-7997-1640]{Santiago Arribas}
\affiliation{Centro de Astrobiolog\'{\i}a (CAB), CSIC-INTA, Ctra. de Ajalvir km 4, Torrej\'on de Ardoz, E-28850, Madrid, Spain}
\author[0009-0005-5448-5239]{Carmen Blanco-Prieto}
\affiliation{Centro de Astrobiolog\'{\i}a (CAB), CSIC-INTA, Ctra. de Ajalvir km 4, Torrej\'on de Ardoz, E-28850, Madrid, Spain}
\author[0000-0002-0984-7713]{Yurina Nakazato}
\affiliation{Department of Physics, The University of Tokyo, 7-3-1 Hongo, Bunkyo, Tokyo 113-0033, Japan}
\author[0000-0001-7925-238X]{Naoki Yoshida}
\affiliation{Department of Physics, The University of Tokyo, 7-3-1 Hongo, Bunkyo, Tokyo 113-0033, Japan}
\affiliation{Kavli Institute for the Physics and Mathematics of the Universe (WPI), UT Institute for Advanced Study, The University of Tokyo, Kashiwa, Chiba 277-8583, Japan}
\affiliation{Research Center for the Early Universe, School of Science, The University of Tokyo, 7-3-1 Hongo, Bunkyo, Tokyo 113-0033, Japan}
\author[0000-0002-5268-2221]{Tom J. L. C. Bakx}
\affiliation{Department of Space, Earth, \& Environment, Chalmers University of
Technology, Chalmersplatsen 4 412 96 Gothenburg, Sweden}
\affiliation{Department of Physics, Graduate School of Science, Nagoya University, Nagoya 464-8602, Japan}
\affiliation{National Astronomical Observatory of Japan, 2-21-1, Osawa, Mitaka, Tokyo, Japan}
\author[0000-0002-8680-248X]{Daniel Ceverino}
\affiliation{Universidad Autonoma de Madrid, Ciudad Universitaria de Cantoblanco, E-28049 Madrid, Spain}
\affiliation{CIAFF, Facultad de Ciencias, Universidad Autonoma de Madrid, E-28049 Madrid, Spain}
\author[0000-0001-6820-0015]{Luca Costantin}
\affiliation{Centro de Astrobiolog\'{\i}a (CAB), CSIC-INTA, Ctra. de Ajalvir km 4, Torrej\'on de Ardoz, E-28850, Madrid, Spain}
\author[0000-0003-2119-277X]{Alejandro Crespo G\'omez}
\affiliation{Space Telescope Science Institute (STScI), 3700 San Martin Drive, Baltimore, MD 21218, USA}
\author[0000-0001-8083-5814]{Masato Hagimoto}
\affiliation{Department of Physics, Graduate School of Science, Nagoya University, Nagoya 464-8602, Japan}
\author[0000-0003-3278-2484]{Hiroshi Matsuo}
\affiliation{National Astronomical Observatory of Japan,
2-21-1 Osawa, Mitaka, Tokyo 181-8588, Japan}
\affiliation{Graduate University for Advanced Studies (SOKENDAI), 2-21-1 Osawa, Mitaka, Tokyo 181-8588, Japan}
\author{Wataru Osone}
\affiliation{Division of Physics, Faculty of Pure and Applied Sciences, University of Tsukuba, Tsukuba, Ibaraki 305-8571, Japan}
\author[0000-0002-6510-5028]{Yi W. Ren}
\affiliation{Department of Pure and Applied Physics, School of Advanced Science and Engineering, Faculty of Science and Engineering, Waseda University, 3-4-1 Okubo, Shinjuku, Tokyo 169-8555, Japan}
\author[0000-0001-7440-8832]{Yoshinobu Fudamoto}
\affiliation{Center for Frontier Science, Chiba University, 1-33 Yayoi-cho, Inage-ku, Chiba 263-8522, Japan}
\author{Takeshi Hashigaya}
\affiliation{Department of Astronomy, Kyoto University Sakyo-ku, Kyoto 606-8502, Japan}
\author[0000-0002-4005-9619]{Miguel Pereira-Santaella}
\affiliation{Instituto de F\'isica Fundamental (IFF), CSIC, Serrano 123, E-28006, Madrid, Spain}
\author[0000-0003-4807-8117]{Yoichi Tamura}
\affiliation{Department of Physics, Graduate School of Science, Nagoya University, Nagoya 464-8602, Japan}

%%%%%%%%%%%%%%%%%%%%%%%%%%%%%%%%%%%%%%%%%%%%%%%%%%%%%%%%%%%%%%%

\begin{abstract}
We report the discovery of a complex, density-stratified interstellar medium (ISM) in the star-forming galaxy COS-2987 at $z = 6.81$, revealed by the unprecedented synergy between JWST/NIRSpec IFS and ALMA observations.
These observations detect key emission lines, including \oii~$\lambda\lambda$~3727, 3730, \oiii~4364, \oiii~$\lambda\lambda$~4960, 5008, \oiii\ 88 \micron, as well as H$\alpha$ and H$\beta$.
JWST spectroscopy alone indicates ISM properties that are typical for galaxies at $z\sim7$. These include low dust extinction ($A_{\rm V} \approx 0.14$ mag), moderate electron density ($n_{\rm e} \approx 500$~cm$^{-3}$), and low gas-phase metallicity ($\sim10\%$). However, the strong far-infrared \oiii\ 88 \micron\ emission detected by ALMA cannot be explained by a single-component ionized medium with uniform electron density and temperature.
Instead, a two-component ISM model, comprising compact, high-temperature, high-density gas components ($T_e \approx 26,000$\,K; $n_e \approx 600~\mathrm{cm}^{-3}$) and an extended, cooler, lower-density component ($T_e \approx 8,000$\,K; $n_e \approx 50~\mathrm{cm}^{-3}$), successfully reproduces the observed line ratios of \oiii~88~\micron/\oiii~5008~\AA\ and \oiii~4364/\oiii~5008~\AA, with a volume ratio of 1 : 300 between the two components. 
Our results demonstrate that JWST alone probes only a fraction of the ISM and highlight the critical importance of combining JWST and ALMA to reveal the density-stratified ISM of early galaxies.
\end{abstract}


%%%%%%%%%%%%%%%%%%%%%%%%%%%%%%%%%%%%%%%%%%%%%%%%%%%%%%%%%%%%%%%
\section{Introduction}\label{sec:intro}
%%%%%%%%%%%%%%%%%%%%%%%%%%%%%%%%%%%%%%%%%%%%%%%%%%%%%%%%%%%%%%%

% general introduction 
Understanding the physical properties of the interstellar medium (ISM) in galaxies is crucial for studying galaxy formation and evolution. With the advent of the \textit{James Webb Space Telescope} (JWST; \citealp{Gardner}) and the Atacama Large Millimeter/submillimeter Array (ALMA; \citealp{Wootten09}), fundamental physical quantities in {\sc H ii} regions, such as electron density ($n_{\rm e}$), electron temperature ($T_{\rm e}$), ionization parameter, and gas-phase metallicity, can now be measured even in galaxies at redshift ($z$) beyond $4$ out to the epoch of reionization ($z \gtrsim 6$) (e.g., \citealp{Killi23, Isobe23, Nakajima23, Arribas24, Laseter24, Alvarez-Marquez25, Abdurro'uf24, Sanders24, zavala25}).

%stratified ISM and the motivation 
Recent studies have reported evidence for a density-stratified ISM in high-$z$ galaxies based on emission-line diagnostics spanning the rest-frame ultraviolet (UV) and optical wavelengths. Specifically, $n_{\rm e}$ derived from high-ionization UV lines such as \ciii~$\lambda\lambda$~1907, 1909 and \niv~$\lambda \lambda$~1483, 1486\footnote{\ciii\ has an ionization potential of 24.4~eV, with critical densities of $\sim 8.7 \times 10^4$ and $\sim 2.8 \times 10^5$~cm$^{-3}$ for $\lambda\lambda$1907, 1909, respectively. \niv\ has a higher ionization potential of 47.5~eV, and $n_{\rm crit}$ values of $\sim 1.5 \times 10^5$ and $\sim 5.0 \times 10^5$~cm$^{-3}$ for $\lambda\lambda$1483, 1486 (e.g., \citealt{Osterbrock06}).}often exceed $10^4$--$10^5$~cm$^{-3}$, while lower densities ($10^2$--$10^3$~cm$^{-3}$) are inferred from optical diagnostics such as \oii~$\lambda\lambda$~3727, 3730 or \sii~$\lambda\lambda$ 6718, 6733\footnote{\oii\ has an ionization potential of 13.6~eV, with critical densities of $\sim 1.4 \times 10^3$ and $\sim 3.4 \times 10^3$~cm$^{-3}$ for $\lambda\lambda$3727, 3730 respectively. \sii\ has an ionization potential of 10.4~eV, and $n_{\rm crit}$ values of $\sim 1.5 \times 10^3$ and $\sim 3.9 \times 10^3$~cm$^{-3}$ for $\lambda\lambda$6718, 6733 (e.g., \citealt{Osterbrock06}).} (e.g., \citealt{Mingozzi22, Ji+Ubler24, Topping25, Alvarez-Marquez25}; A. Crespo G\'omez et al., in preparation).
This apparent discrepancy is interpreted as evidence for a density-stratified ISM, in which UV lines such as \ciii\ and \niv\ trace highly ionized, high-density regions, whereas optical lines like \oii\ and \sii\ originate from lower-ionization, lower-density gas.

A similar trend is also seen in the far-infrared regime. Based on {\it Herschel}/PACS observations of LMC-N11, the second largest giant H\,{\sc ii} region in the LMC after 30 Doradus, \citet{Lebouteiller12} found that \oiii~88~\micron\ emission is spatially extended well beyond the area predicted from the number of massive stars. The emission requires low-density gas with $n_{\rm e} < 16$~cm$^{-3}$, which is much lower than the values inferred from optical \oii\ and \sii\ doublets ($\sim 100$~cm$^{-3}$) or \cliii\ $\lambda\lambda$5517, 5538 ($\sim 1700$~cm$^{-3}$). They interpreted this as evidence that \oiii~88~\micron\ emission is photoionized by far-UV photons that penetrate into low-density regions due to a clumpy ISM structure.

Motivated by these findings, we investigate whether a similar stratification of highly ionized gas can be identified by combining rest-frame optical and far-infrared (FIR) \oiii\ emission lines\footnote{\oiii\ has an ionization potential of 35.1~eV, and critical densities of $3.0 \times 10^6$, $6.8 \times 10^5$, and 510~cm$^{-3}$ for $\lambda$4364~\AA, $\lambda$5008~\AA, and 88~\micron, respectively (e.g., \citealt{Osterbrock06}).}.

%optical and FIR [OIII] emission lines 
In principle, one can determine both $T_{\rm e}$ and $n_{\rm e}$ of an {\sc H ii} region with uniform physical conditions using emission lines of \oiii.
For example, the flux ratio of [O\,\textsc{iii}]~$5008$~\AA/$4364$~\AA\ depends primarily on $T_{\rm e}$, whereas the flux ratio of [O\,\textsc{iii}]~88~$\mu$m/$5008$~\AA\ depends on both $T_{\rm e}$ and $n_{\rm e}$ (e.g., \citealt{Dinerstein85,Osterbrock06,Kewley19,Jones20,Yang20,Nakazato23}). 
This technique has recently been applied to local analogues of high-$z$ galaxies to investigate ISM properties such as $T_{\rm e}$, $n_{\rm e}$, and elemental abundance ratios (e.g., \citealp{Chen23, Chen24}; see also \citealp{Kumari24}).

At $z \gtrsim 6$, the powerful synergy between JWST and ALMA enables, for the first time, complementary observations of rest-frame optical and FIR [O\,\textsc{iii}] emission lines. \cite{Fujimoto24}, \cite{Stiavelli23}, and \cite{Harshan24} have estimated $n_{\rm e}$ in $z \sim 8$--$9$ galaxies based on measurements of $T_{\rm e}$ from optical spectroscopy and the flux ratio of [O\,\textsc{iii}]~88~$\mu$m/$5008$~\AA.
However, these studies utilized the NIRSpec multi-object spectroscopy (MOS) mode to observe optical [O\,\textsc{iii}] emission lines, which are subject to uncertainties arising from slit-loss corrections and from effects related to the position angle, particularly when the high-$z$ source has a clumpy morphology or is elongated along a specific axis. In contrast, ALMA observations are free from such observational limitations.
To accurately characterize the global properties of galaxies, a combination of NIRSpec integral field spectroscopy (IFS; \citealp{Boker22}) and ALMA observations is a crucial next step (see e.g., \citealp{Scholtz25, zavala25}).

%RIOJA and this study 
The RIOJA (Reionization and the ISM/Stellar Origins with JWST and ALMA) project (JWST GO1 PID 1840; PIs: J. \'{A}lvarez-M\'{a}rquez and T. Hashimoto; \citealp{Hashimoto23, Sugahara25,Mawatari25}) conducts follow-up observations of 12 ALMA [O\,\textsc{iii}]~$88~\mu$m emitters using NIRSpec/IFS and NIRCam.
Here, we present a combined analysis of optical and FIR [O\,\textsc{iii}] emission lines in \name\ (hereafter \shortname; \citealp{Smit15, Laporte17, Smit18, Witstok22, Posses23, Harikane25a,Mawatari25}), a star-forming galaxy at $z = 6.81$. 
%\footnote{This paper is based on the master's thesis of the first author, which was approved in March 2025 \citep{Usui2025Master}, and on an oral presentation given at the 2024 Spring Meeting of the Astronomical Society of Japan \citep{Usui24asj}.}.
%
We find that a simple homogeneous ISM model cannot reproduce the observed \oiii\ line ratios, suggesting the presence of hidden complexity in the ionized gas structure, such as multiple ionized gas components with distinct physical conditions in terms of $T_{\rm e}$ and $n_{\rm e}$. 
%\footnote{A similar interpretation has been reached independently by \citet{Harikane25b}, who conducted a joint analysis of JWST and ALMA data for a sample of galaxies at $z = 6$--$9$.}.

In a companion paper (\citealt{Mawatari25}), we provide an overview of the observations of \shortname, discuss its stellar populations based on combined JWST and ALMA data, and present the ISM properties derived from JWST's spatially-resolved observations. 

%definitions 
Throughout this paper, we adopt a $\Lambda$CDM cosmology with $\Omega_{m} = 0.3$, $\Omega_{\Lambda} = 0.7$, and $H_{0} = 70$\,km\,s$^{-1}$\,Mpc$^{-1}$. We assume a solar luminosity of $L_{\odot} = 3.839 \times 10^{33}$\,erg\,s$^{-1}$ and a solar metallicity of $12+\log(\mathrm{O/H})_{\odot} = 8.69$ (\citealp{Asplund09}).
Hereafter, \oii~$\lambda \lambda$~3727, 3730, \oiii~4364~\AA, \oiii~$\lambda \lambda$~4960, 5008 and \oiii~88~$\mu$m are referred to as \oii3727, 3730, \oiii4364, \oiii4960, 5008 and \oiii88, respectively, unless otherwise specified.

%%%%%%%%%%%%%%%%%%%%%%%%%%%%%%%%%%%%%%%%%%%%%%%%%%%%%%%%%%%%%%%
\section{Observations and Data}\label{sec:data}
%%%%%%%%%%%%%%%%%%%%%%%%%%%%%%%%%%%%%%%%%%%%%%%%%%%%%%%%%%%%%%%

%%%%%%%%%%%%%%%%%%%%%%%%%%%%%%%%%%%%%%%%%%%%%%%%%%%%%%%%%%%%%%%
\subsection{JWST NIRSpec IFS Data}\label{subsec:data1}
%%%%%%%%%%%%%%%%%%%%%%%%%%%%%%%%%%%%%%%%%%%%%%%%%%%%%%%%%%%%%%%

The JWST NIRSpec IFS data were obtained as part of the RIOJA project. The specific observations can be accessed via the Mikulski Archive for Space Telescopes (MAST) at the Space Telescope Science Institute:
\dataset[doi:10.17909/6mmv-2944]{https://doi.org/10.17909/6mmv-2944}.
Details of the observations and data reduction procedures are presented in \cite{Mawatari25}. Briefly, we used the G395H/F290LP grating/filter combination, which provides a high spectral resolution ($R \sim 2700$) and covers a wavelength range from 2.87 to 5.27\,$\mu$m.

The raw data were processed using the JWST reduction pipeline version 1.14.0 (\citealp{Bushouse24}) under the CRDS context \texttt{jwst\_1223.pmap}.
We performed the following custom steps to improve the data quality: (1) sigma clipping to reject bad pixels and cosmic rays, and (2) subtraction of the median background from the calibrated images (e.g., \citealp{Ubler23, Perna23, Marshall23}). The final data cube was combined using the ``drizzle'' method, sampled with a pixel size of $0\farcs05$. 
To ensure that the line fluxes are measured from the same spatial region, the point-spread function (PSF) of the cube was matched to that of the H$\alpha$ emission line, which corresponds to the longest wavelength among the detected lines (see Appendix \ref{appendix1}).

\begin{figure*}[t]
    \begin{center}
        \includegraphics[width=16cm]{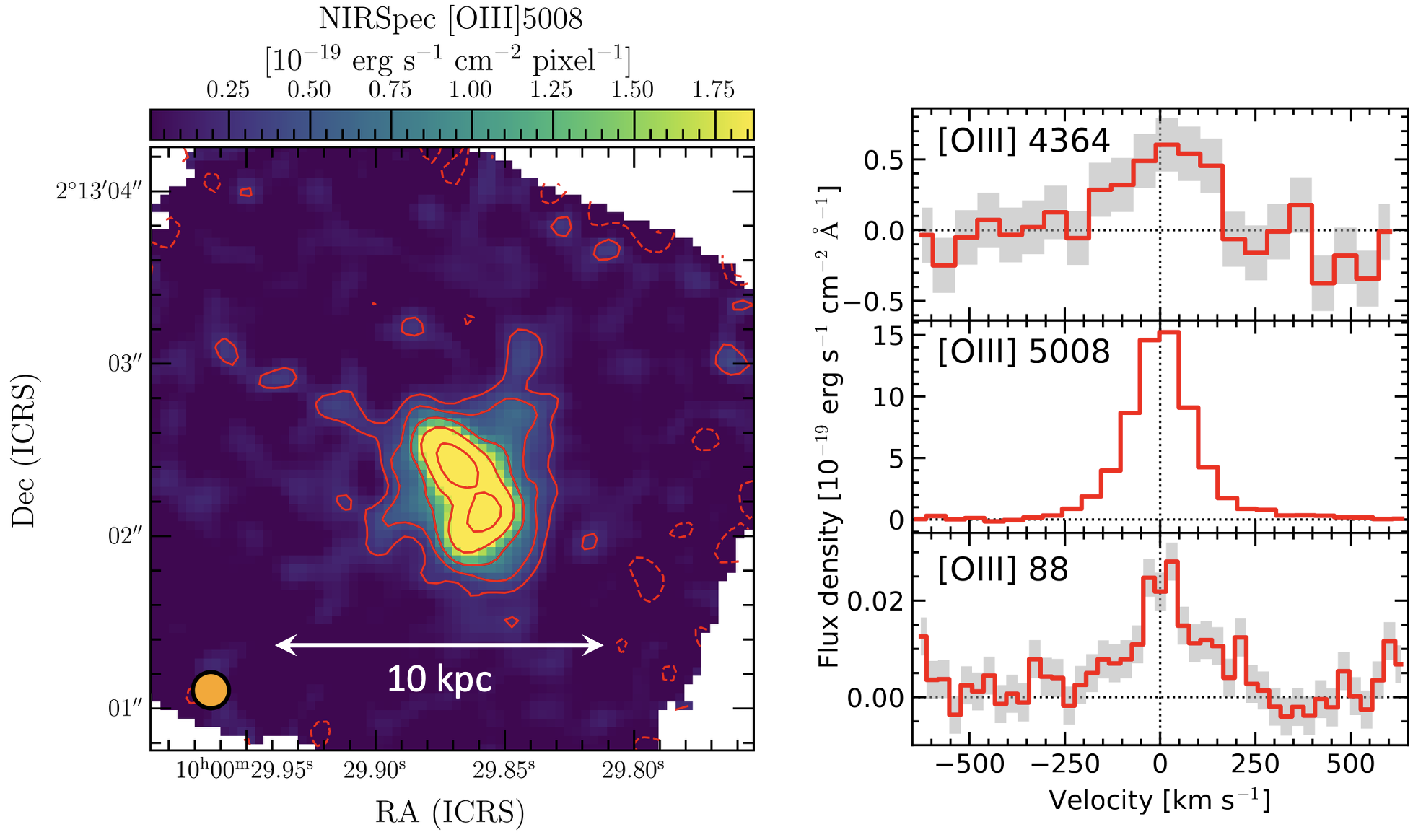}\\
    \end{center}
    \vspace{-0.5cm}
    \caption{
    (\textit{Left}) Integrated intensity map of the NIRSpec \oiii5008 emission. The red contours show the $\pm2^{n}\sigma$ significance levels ($n = 1, 2, 3, \dots$), where $\sigma = 1.16 \times 10^{-20}\,{\rm erg\,s^{-1}\,cm^{-2}\,pixel^{-1}}$. Positive and negative contours are shown by the solid and dashed lines, respectively. The orange circle at the bottom left indicates the FWHM of $0\farcs21$ at the observed wavelength of H$\alpha$ ($5.125$\,$\mu$m).
    (\textit{Right}) One-dimensional spectra of \oiii\,4364, \oiii\,5008, and \oiii\,88, where the horizontal axis is the velocity relative to the best-fit redshift to the \oiii\,5008 line. In each panel, the red solid histogram and gray shade indicate the observed flux density and $1\sigma$ uncertainty, respectively. The gray shade is not visible for \oiii5008 due to its negligible $1\sigma$ uncertainty.
    }
    \label{fig:fig1_OIII_3lines}
\end{figure*}

As presented in \cite{Mawatari25}, we detected multiple emission lines, including \oii3727, 3730, H$\beta$, \oiii4364, \oiii4960, 5008, and H$\alpha$, all at $\gtrsim 4\sigma$ significance in the integrated spectrum.
The left panel of Figure~\ref{fig:fig1_OIII_3lines} shows the integrated intensity map of the \oiii5008 emission, while the spatially integrated one-dimentional spectra of \oiii4364 and \oiii5008 are shown in the right panel.

Table~\ref{tab:SpatiallyIntegratedProp} shows the line fluxes.
The fluxes were obtained by applying Gaussian fitting to each line profile, where the spectra were extracted from regions with $\gtrsim 2\sigma$ significance in the \oiii5008 intensity map (Figure \ref{fig:fig1_OIII_3lines}). Based on a curve-of-growth analysis, we estimate that the adopted aperture misses $\approx$~10\% of the total flux. Therefore, we applied a 10\% aperture correction to all line flux values, because we use the PSF-matched data. This aperture correction does not affect the line ratios of NIRSpec IFS data, but is required when comparing NIRSpec IFS and ALMA data. 

%FFFFFFFFFFFFFFFFFFFFFFFFFFFFFFFFFFFFFFFFFFFFFFFFFFFFFFFFFFFFF%
%FFFFFFFFFFFFFFFFFFFFFFFFFFFFFFFFFFFFFFFFFFFFFFFFFFFFFFFFFFFFF%

%%%%%%%%%%%%%%%%%%%%%%%%%%%%%%%%%%%%%%%%%%%%%%%%%%%%%%%%%%%%%%%
\subsection{\texorpdfstring{ALMA \oiii\ 88 \micron\ data}{ALMA [OIII] 88 um data}}\label{subsec:data2}
%%%%%%%%%%%%%%%%%%%%%%%%%%%%%%%%%%%%%%%%%%%%%%%%%%%%%%%%%%%%%%%

ALMA Band 8 (\citealt{Sekimoto08})\footnote{See also \cite{Bakx24} for a summary of ALMA receiver references.} observations targeting \oiii88 in \shortname\ were conducted as part of an ALMA Cycle 6 program (ID: 2018.1.00429.S, PI: R. Smit). Although \cite{Witstok22} presented the initial detection and results of the \oiii88 line, we reanalyzed the data cube to uniformly measure the line flux of \shortname\ in a manner consistent with other RIOJA studies.

The data were reduced using the Common Astronomy Software Applications (CASA; \citealt{mcmullin2022}) version 5.6. Imaging was performed with the CASA \texttt{tclean} task, adopting natural weighting to optimize point-source sensitivity. The final cube has a synthesized beam size of $0\farcs80 \times 0\farcs61$ with a positional angle of $86^\circ$, consistent with \cite{Witstok22}. The typical rms noise of the cube is $0.576$\,mJy\,beam$^{-1}$ per 30\,km\,s$^{-1}$ channel.

As shown in Figure \ref{apdx_fig:OIII_SpatialDistribution} in Appendix \ref{appendix2}, we integrated the emission over the frequency range of 434.2--434.9\,GHz, corresponding to a velocity width of 480\,km\,s$^{-1}$ (i.e., $\approx3$~times the FWHM of the optical lines), to generate the integrated intensity (moment-0) map of the \oiii88 emission. The noise level of the map is $\sigma = 82.6$\,mJy\,beam$^{-1}$\,km\,s$^{-1}$, and the \oiii88 line is detected at a peak significance of $7.5\sigma$.
We measure an integrated line flux of $1.06 \pm 0.23$\,Jy\,km\,s$^{-1}$, corresponding to a flux of $(15.3 \pm 3.3) \times 10^{-18}$\,erg\,s$^{-1}$\,cm$^{-2}$, consistent with the values reported in \cite{Witstok22} within uncertainties. The bottom right panel of Figure \ref{fig:fig1_OIII_3lines} shows the spectrum of \oiii88 emission.

%ttttttttttttttttttttttttttttttttttttttttttttttttttttttttttttt%
\begin{deluxetable}{lccc}
    \tablecaption{The spatially-integrated properties for \shortname.\label{tab:SpatiallyIntegratedProp}}
    \tabletypesize{\small} 
    \tablewidth{0pt}
    \tablecolumns{4}
    \tablehead{Line fluxes\tablenotemark{{\rm \dag}} [10$^{-18}$ erg s$^{-1}$ cm$^{-2}$]}
    \startdata
        \hline\hline
        $F_{\rm [OII]3727}$  & & & $7.1 \pm 0.9$ \\ 
        $F_{\rm [OII]3730}$  & & & $7.6 \pm 1.2$ \\ 
        $F_{\rm [OIII]4364}$  & & &  $1.8 \pm 0.5$ \\ 
        $F_{\rm H\beta}$  & & & $8.0 \pm 0.8$ \\ 
        $F_{\rm [OIII]4960}$  & & & $17.4 \pm 1.8$ \\ 
        $F_{\rm [OIII]5008}$  & & & $52.3 \pm 5.4$ \\ 
        $F_{\rm H\alpha}$  & & & $23.9 \pm 2.4$ \\  
        $F_{\rm [OIII]88}$  & & & $15.3 \pm 3.3$ \\ 
        \hline\hline
        \multicolumn{4}{l}{Line Ratios\tablenotemark{{\rm \dag \dag}}} \\
        \hline\hline
        \ha/\hb           & & & $3.00\pm0.40$ \\
        \oiii4364/5008                               & & & $0.038_{-0.011}^{+0.010}$\\ 
        \oii3730/3727                                & & & $1.05\pm0.21$ \\
        \oii3727,3730/\hb                            & & & $1.99_{-0.37}^{+0.41}$ \\
        \oiii4960,5008/\hb                           & & & $8.64_{-0.09}^{+0.13}$ \\
        \oiii88/5008$^{a}$   & & & $0.22_{-0.07}^{+0.09}$ \\ 
        \hline\hline
        \multicolumn{4}{l}{Physical properties} \\
        \hline\hline
        \av/mag                             & & & $0.14_{-0.14}^{+0.32}$\\
        \temp(\oiii)/K                      & & & $2.02_{-0.28}^{+0.30} \times 10^4$\\
        \temp(\oii)/K$^{b}$                        & & & $1.72_{-0.20}^{+0.21} \times 10^4$\\
        \dens(\oii)/${\rm cm^{-3}}$         & & & $494_{-300}^{+546}$ \\
        12+log(O/H)                         & & & $7.67^{+0.11}_{-0.13}$ \\
        \hline\hline
        \multicolumn{4}{l}{Dust-corrected line luminosity [$\times10^8$~\lsun]} \\
        \hline\hline
        $L_{{\rm [O~\textsc{iii}]4364}}$ & & & $3.35_{-1.17}^{+2.08}$ \\
        $L_{{\rm [O~\textsc{iii}]5008}}$ & & & $88.5_{-17.2}^{+44.2}$ \\
        $L_{{\rm [O~\textsc{iii}]88}}$   & & & $21.4\pm4.6$ \\
    \enddata
    \tablenotetext{\dag}{All values before the dust correction.}
    \tablenotetext{\dag \dag}{All values corrected by dust extinction, except the Balmer decrement, for which we present the observed value.}
    \tablenotetext{^{a}}{The \oiii88/5008 ratio includes 10\% absolute flux calibration uncertainties for both \oiii5008 and 88 fluxes (see \S\ \ref{sec:result_o88}).}
    \tablenotetext{^{b}}{Scaled from \temp(\oiii) with eq. (\ref{eq:te_converted}).}
\end{deluxetable}
%ttttttttttttttttttttttttttttttttttttttttttttttttttttttttttttt%

%%%%%%%%%%%%%%%%%%%%%%%%%%%%%%%%%%%%%%%%%%%%%%%%%%%%%%%%%%%%%%%
\section{Properties Obtained from NIRSpec/IFS}\label{sec:result}
%%%%%%%%%%%%%%%%%%%%%%%%%%%%%%%%%%%%%%%%%%%%%%%%%%%%%%%%%%%%%%%

%%%%%%%%%%%%%%%%%%%%%%%%%%%%%%%%%%%%%%%%%%%%%%%%%%%%%%%%%%%%%%%
\subsection{Integrated Internal Dust Extinction}\label{subsubsec:result_Av}
%%%%%%%%%%%%%%%%%%%%%%%%%%%%%%%%%%%%%%%%%%%%%%%%%%%%%%%%%%%%%%%

We estimate the internal dust extinction (\av) of \shortname\ to correct the observed emission line fluxes. We derive \av\ based on the Balmer decrement of H$\alpha$/H$\beta$ as follows\footnote{At the sky position of the target and over the wavelength range of $\sim3$--$5$\,$\mu$m, the Galactic extinction is negligible ($A_{\lambda} < 0.01$\,mag; \citealt{Schlegel98}).}:
\begin{equation}
    A_{V} = \frac{2.5 \times R_{V}}{k(\lambda_{\rm H\beta}) - k(\lambda_{\rm H\alpha})} \log_{10} \left[ \frac{({\rm H\alpha/H\beta})_{\rm obs}}{({\rm H\alpha/H\beta})_{\rm theo}} \right],
    \label{eq:AV}
\end{equation}
where $k(\lambda)$ denotes the extinction curve, and $R_V \equiv A_V/E(B-V)$ is the total-to-selective extinction ratio. We adopt the SMC extinction curve with $R_V=2.74$ (\citealp{Gordon03}). $({\rm H\alpha/H\beta})_{\rm obs}$ and $({\rm H\alpha/H\beta})_{\rm theo}$ represent the observed and theoretical Balmer line ratios, respectively. 
We obtain $({\rm H\alpha/H\beta})_{\rm obs} = 3.00\pm0.40$ (Table~\ref{tab:SpatiallyIntegratedProp}).

To account for the dependence of $({\rm H\alpha/H\beta})_{\rm theo}$ on electron density and temperature, we perform an iterative calculation following \cite{Welch24}, who carefully derived ISM properties for two high-$z$ star-forming galaxies using data from the TEMPLATES JWST Early Release Science (ERS) program (\citealp{Rigby25})\footnote{
We first calculate $({\rm H\alpha/H\beta})_{\rm theo}$ in Case B recombination, assuming an electron density of $n_{\rm e} = 100\,{\rm cm^{-3}}$ and an electron temperature of $T_{\rm e} = 10^4$\,K, using the {\tt PyNeb RecAtom} function (\citealp{Luridiana15}). We then derive \av\ from eq.~(\ref{eq:AV}), correct the observed emission line fluxes for dust attenuation, and remeasure the electron density and temperature from the \oii3730/3727 and \oiii5008/4364 line ratios, respectively. This process is repeated until the change in temperature between iterations becomes less than 10\,K.}.
To estimate the uncertainty in \av, we perform a Monte Carlo simulation with 10,000 realizations, randomly perturbing the observed H$\alpha$ and H$\beta$ fluxes within their uncertainties.

We find \av\,$= 0.14^{+0.32}_{-0.14}$\,mag. This result indicates that \shortname\ is dust-poor, consistent with the non-detection of dust continuum emission by ALMA (\citealp{Smit18, Witstok22}). Hereafter, unless otherwise stated, we base our analysis on the dust-corrected line ratios in Table~\ref{tab:SpatiallyIntegratedProp}.

%%%%%%%%%%%%%%%%%%%%%%%%%%%%%%%%%%%%%%%%%%%%%%%%%%%%%%%%%%%%%%%
\subsection{Electron Temperature}\label{subsubsec:result_Te}
%%%%%%%%%%%%%%%%%%%%%%%%%%%%%%%%%%%%%%%%%%%%%%%%%%%%%%%%%%%%%%%

We measure the electron temperature of the highly ionized gas, $T_{\rm e}$(\oiii), based on the \oiii4364/5008 line ratio using the {\tt PyNeb getTemDen} function (\citealp{Luridiana15}). The uncertainty in $T_{\rm e}$(\oiii) is estimated via a Monte Carlo method, following the same approach as for $A_V$ (\S\ref{subsubsec:result_Av}). Based on the measured line ratio of \oiii4364/5008\,$=0.038_{-0.011}^{+0.010}$ (Table~\ref{tab:SpatiallyIntegratedProp}), we derive an electron temperature of $T_{\rm e}$(\oiii)\,$=2.02^{+0.30}_{-0.28} \times 10^4$\,K. This value is comparable to those found in other $z \sim 7$ galaxies with similar gas-phase metallicity to \shortname\ ($\approx10\%$ of the solar metallicity; see \S~\ref{subsubsec:result_metallicity})z observed by JWST (e.g., \citealp{Schaerer22, Rhoads23, Nakajima23, Sanders24, Laseter24, Morishita24, Curti24}).

We estimate the electron temperature of the lower ionization gas, $T_{\rm e}$(\oii), using the relation from \cite{Campbell1986, Stasinska1982}:
\begin{equation}
T_{\rm e}({\rm [O\,\textsc{ii}]}) = 0.7 \times T_{\rm e}({\rm [O\,\textsc{iii}]}) + 3000\ {\rm K}.
\label{eq:te_converted}
\end{equation}
Applying this relation, we obtain $T_{\rm e}$(\oii)\,$=1.72^{+0.21}_{-0.20} \times 10^4$\,K.

%%%%%%%%%%%%%%%%%%%%%%%%%%%%%%%%%%%%%%%%%%%%%%%%%%%%%%%%%%%%%%%
\subsection{Electron Density}
%%%%%%%%%%%%%%%%%%%%%%%%%%%%%%%%%%%%%%%%%%%%%%%%%%%%%%%%%%%%%%%

%FFFFFFFFFFFFFFFFFFFFFFFFFFFFFFFFFFFFFFFFFFFFFFFFFFFFFFFFFFFFF%
\label{subsubsec:result_ne}
\begin{figure}[t]
    \begin{center}
        \includegraphics[scale=0.55]{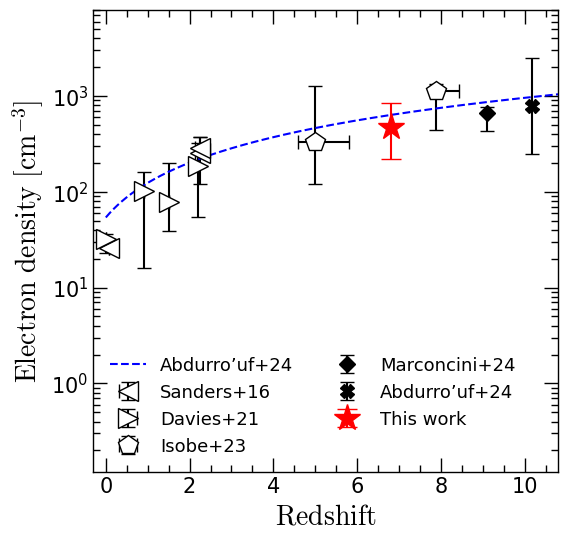}\\
    \end{center}
    \vspace{-0.5cm}
    \caption{
    Redshift evolution of the electron density $n_{\rm e}$. The red star indicates the electron density of \shortname, derived from the \oii\ ratio. 
    Literature data at $z \sim 0$--3 are compiled from \cite{Sanders16} (\sii\ or \oii) and \cite{Davies21} (\sii), while data at $z \gtrsim 4$ are taken from \cite{Isobe23}, \cite{Abdurro'uf24}, and \cite{Marconcini24}, all based on the \oii\ ratios. Filled symbols represent measurements for individual galaxies, and open symbols represent median values for the sample galaxies. The electron density of \shortname, derived from the \oii\ ratio, is consistent with the overall redshift evolution trend indicated by the blue dashed line (\citealt{Abdurro'uf24}).
    }
    \label{fig:Fig2_z_vs_ne}
\end{figure}
%FFFFFFFFFFFFFFFFFFFFFFFFFFFFFFFFFFFFFFFFFFFFFFFFFFFFFFFFFFFFF%

We estimate the electron density based on the \oii3730/3727 line ratio, using the {\tt PyNeb getTemDen} function (\citealp{Luridiana15}). We adopt $T_{\rm e}$(\oii) derived in \S\ref{subsubsec:result_Te}, although the \oii3730/3727 ratio has a weak dependence on electron temperature. The uncertainty in $n_{\rm e}$(\oii) is evaluated using a Monte Carlo method, following the same procedure as for $A_V$ (\S\ref{subsubsec:result_Av}). From the measured line ratio of \oii3730/3727\,$=1.05\pm0.21$ (Table~\ref{tab:SpatiallyIntegratedProp}), we derive an electron density of $n_{\rm e}$(\oii)\,$=494^{+546}_{-300}\,{\rm cm^{-3}}$.

Figure~\ref{fig:Fig2_z_vs_ne} shows the redshift evolution of electron density, where the red star represents $n_{\rm e}$ of \shortname. The $n_{\rm e}$ values at $z \sim 0$--$3$ in \cite{Davies21} are derived from the \sii\ ratio, while those in \cite{Sanders16} are obtained from either \sii\ or \oii\ line ratios.
Our measurement is consistent with typical values for $z\sim5-10$ galaxies derived from the \oii\ ratio (\citealp{Isobe23, Abdurro'uf24, Marconcini24}), and also agrees with the best-fit redshift--electron density relation shown by the blue dashed line (\citealp{Abdurro'uf24}). We stress that the data point of \shortname\ nicely fills in the redshift gap ($z\sim6-8$) in the previous study of \cite{Isobe23}.

%%%%%%%%%%%%%%%%%%%%%%%%%%%%%%%%%%%%%%%%%%%%%%%%%%%%%%%%%%%%%%%
\subsection{Gas-Phase Metallicity based on the direct-$T_{\rm e}$ method}\label{subsubsec:result_metallicity}
%%%%%%%%%%%%%%%%%%%%%%%%%%%%%%%%%%%%%%%%%%%%%%%%%%%%%%%%%%%%%%%

We estimate the oxygen abundance, $12+\log({\rm O/H})$, as a proxy for gas-phase metallicity using the direct-$T_{\rm e}$ method with the {\tt PyNeb getIonAbundance} function (\citealp{Luridiana15}). We assume that the total oxygen abundance O/H is the sum of $\rm{O^{++}/H^{+}}$ and $\rm{O^{+}/H^{+}}$. The $\rm{O^{++}/H^{+}}$ abundance is estimated from the \oiii4960,5008/H$\beta$ ratio and $T_{\rm e}$(\oiii), while the $\rm{O^{+}/H^{+}}$ abundance is derived from the \oii3727,3730/H$\beta$ ratio and $T_{\rm e}$(\oii) (Table~\ref{tab:SpatiallyIntegratedProp}).

The uncertainty in $12+\log({\rm O/H})$ is evaluated using the same Monte Carlo technique adopted for $A_V$ (\S\ref{subsubsec:result_Av}). We obtain a metallicity of $12+\log({\rm O/H})=7.67^{+0.11}_{-0.13}$, corresponding to $10\pm2$\% of the solar metallicity.

%%%%%%%%%%%%%%%%%%%%%%%%%%%%%%%%%%%%%%%%%%%%%%%%%%%%%%%%%%%%%%%
\section{\oiii\ Line Ratio Diagnostics Combining JWST and ALMA}\label{sec:result_o88}
%%%%%%%%%%%%%%%%%%%%%%%%%%%%%%%%%%%%%%%%%%%%%%%%%%%%%%%%%%%%%%%

%FFFFFFFFFFFFFFFFFFFFFFFFFFFFFFFFFFFFFFFFFFFFFFFFFFFFFFFFFFFFF%
\begin{figure*}[t]
    \begin{center}
        \includegraphics[width=0.98\linewidth, angle=0]{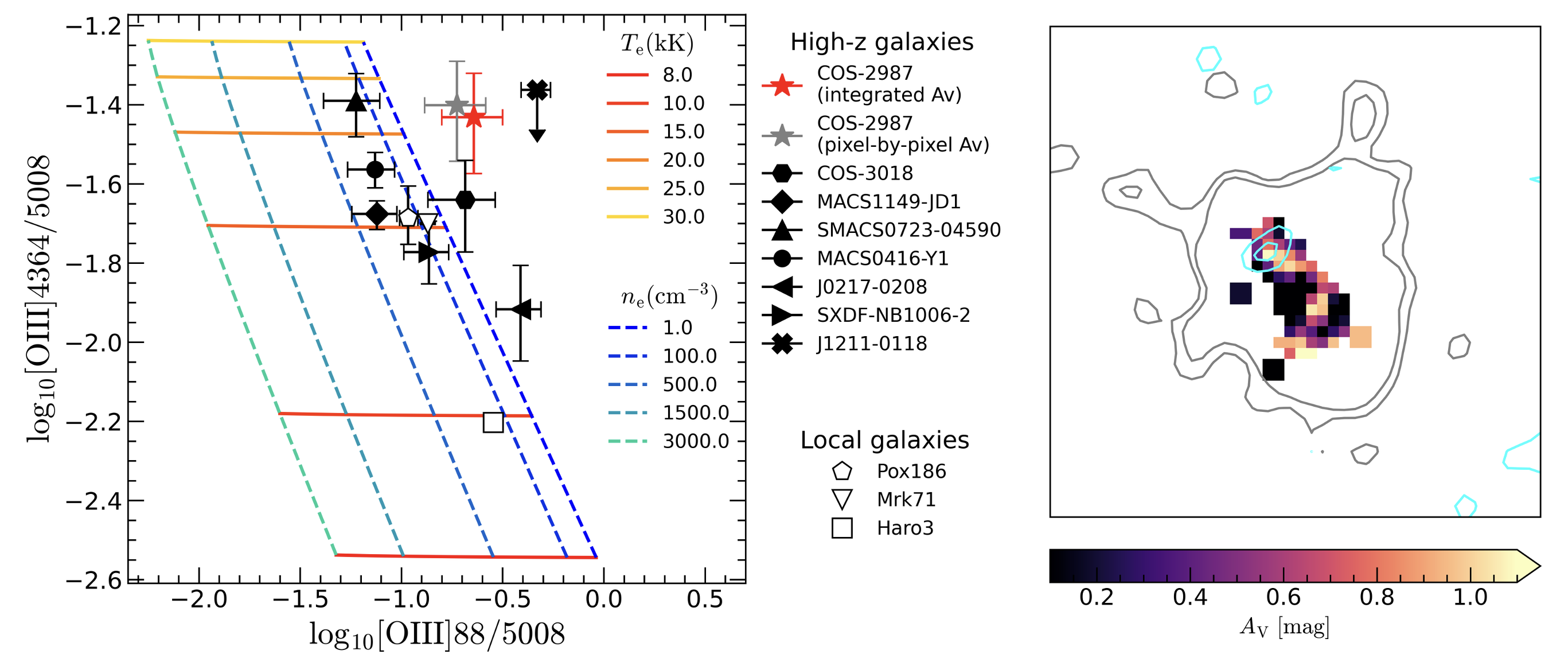}\\
    \end{center}
    \caption{
    (\textit{Left}) Diagnostic diagram of the \oiii\ line ratios overlaid with $T_{\rm e}$--$n_{\rm e}$ model grids generated using {\tt PyNeb} (\citealp{Luridiana15}). The red star indicates the observed \oiii\ line ratios of \shortname\  (Table~\ref{tab:SpatiallyIntegratedProp}). The gray star shows the ratios dust-corrected on a pixel-by-pixel basis using the $A_V$ map shown in the right panel (\S~\ref{subsec:discussion_dust}). Black filled symbols represent high-$z$ star-forming galaxies (\citealp{Witstok22, Stiavelli23, Fujimoto24, Harshan24, Harikane25b}), where the optical \oiii\ data are obtained from NIRSpec/IFS for COS-2987, COS-3018, SXDF-NB1006-2, J0217-0208, and J1211-0118, and from NIRSpec/MOS for the remaining galaxies.
    Open symbols represent local dwarf galaxies (\citealp{Chen23, Kumari24, Chen24}). The \oiii\ line ratios of \shortname\ lie outside the model grids, indicating that they cannot be reproduced by a homogeneous ionized gas with a single set of electron temperature and density.
    (\textit{Right}) $A_V$ map derived from the H$\alpha$/H$\beta$ ratio, constructed within the $3\sigma$ regions of both H$\alpha$ and H$\beta$ integrated intensity maps after the Voronoi binning \citep{CappellariCopin03,Cappellari09}. The gray and cyan contours indicate the \oiii5008 and \oiii4364 integrated intensity maps, respectively, at the $2\sigma$ and $3\sigma$ levels.
    }
    \label{fig:Fig3_OIII_Diagnostics}
\end{figure*}
%FFFFFFFFFFFFFFFFFFFFFFFFFFFFFFFFFFFFFFFFFFFFFFFFFFFFFFFFFFFFF%

One of the major strengths of this study is that we have four \oiii\ emission lines: rest-frame optical emission lines of \oiii4364, \oiii 4960, 5008 observed with JWST/NIRSpec IFS, together with the rest-frame FIR emission line \oiii88 observed with ALMA. Since these are emitted from the same ion of oxygen, we can derive the physical properties of nebulae without assumptions of abundance ratios.

The left panel of Figure~\ref{fig:Fig3_OIII_Diagnostics} presents a diagnostic diagram of the optical \oiii\ line ratio (\oiii4364/5008) versus the FIR-to-optical \oiii\ line ratio (\oiii88/5008), overlaid with model grids covering a wide range of $n_{\rm e}$ and $T_{\rm e}$ values. These model grids are calculated with \texttt{PyNeb} assuming that all \oiii\ lines originate from a homogeneous ionized gas with uniform electron temperature and density.

The red star in the left panel of Figure~\ref{fig:Fig3_OIII_Diagnostics} shows the observed \oiii\ line ratios for \shortname, where the \oiii88/5008 ratio incorporates a 10\% absolute flux calibration uncertainty for both \oiii5008 (\citealp{Boker23}\footnote{See also ``Data Calibration Considerations" in the JWST User Documentation: \url{https://jwst-docs.stsci.edu/depreciated-jdox-articles/jwst-data-calibration-considerations}}) and \oiii88 (ALMA Cycle 6 Proposer's Guide\footnote{\url{https://almascience.eso.org/documents-and-tools/cycle6/alma-proposers-guide}}).

For comparison, we also plot \oiii\ line ratios from the literature. Black filled symbols represent high-$z$ star-forming galaxies. 
These include COS-3018 at $z=6.85$ (\citealt{Smit18, Witstok22, Scholtz25}), SXDF-NB1006-2 at $z=7.21$ (\citealt{Inoue16, Ren23, Harikane25b}; Y. W. Ren et al. 2025, MNRAS, submitted), J0217-0208 at $z=6.20$ (\citealt{Harikane20, Harikane25b}), and J1211-0118 at $z=6.03$ (\citealt{Harikane20, Harikane25b}), where \oiii4364 and \oiii5008 were observed using the NIRSpec/IFS mode and \oiii88 with ALMA. We additionally plot MACS0416-Y1 at $z=8.38$ (\citealt{Bakx20, Tamura23, Harshan24, Hagimoto25}), SMACS0723-04590 at $z = 8.50$ (\citealp{Fujimoto24}) and MACS1149-JD1 at $z = 9.11$ (\citealp{Hashimoto18, Tokuoka22, Stiavelli23}), where \oiii4364 and \oiii5008 were observed using the NIRSpec/MOS\footnote{In MACS0416-Y1, the NIRSpec prism mode do not spectrally resolve \oiii4364 and H$\gamma$, but the authors were able to decompose these line fluxes based on the Balmer decrement obtained from H$\alpha$ and H$\beta$.} mode and \oiii88 with ALMA. 
Black open symbols show local dwarf galaxies, including Mrk\,71 at a distance of 3.4\,Mpc (\citealp{Chen23}), Pox\,186 at $z=0.0041$ (\citealp{Kumari24}), and Haro\,3 at $z=0.0032$ (\citealp{Chen24}), where both the optical and FIR \oiii\ lines were observed with IFS.

In the left panel of Figure~\ref{fig:Fig3_OIII_Diagnostics}, 
the observed ratios for \shortname, COS-3018, and J0217-0218 lie outside the $T_{\rm e}$--$n_{\rm e}$ model grids, whereas the other high-$z$ galaxies fall within the grid.
Moreover, the high-$z$ data points from the literature (black filled symbols) suggest $n_{\rm e} = 100-500$~cm$^{-3}$ while the redshift dependence presented in Figure~\ref{fig:Fig2_z_vs_ne} indicates typical densities of $n_{\rm e} > 500$~cm$^{-3}$. Therefore, the lower-$n_{\rm e}$ values from the FIR-to-optical \oiii\ line ratios seem to be common in high-$z$ galaxies (see also a recent similar conclusion by \citealt{Harikane25b}). 

We stress that the lower $n_{\rm e}$ values derived from the FIR-to-optical \oiii\ ratios are not limited to high-$z$ galaxies. As noted in \S~\ref{sec:intro}, observations of the nearby massive star-forming region LMC-N11 reveal spatially extended and bright \oiii88\ emission, which is inconsistent with the $n_{\rm e}$ values inferred from the optical \oii, \sii, or \cliii\ lines (\citealt{Lebouteiller12}). These optical diagnostics have higher critical densities and therefore likely trace more compact regions than \oiii88.

The results indicate that the \oiii\ line ratios of \shortname\ cannot be reproduced by a simple model of homogeneous ionized gas with uniform temperature and density. 
In \S\ref{sec:discussion}, we discuss two possible origins for this discrepancy.

%%%%%%%%%%%%%%%%%%%%%%%%%%%%%%%%%%%%%%%%%%%%%%%%%%%%%%%%%%%%%%%
\section{Discussion} \label{sec:discussion}
%%%%%%%%%%%%%%%%%%%%%%%%%%%%%%%%%%%%%%%%%%%%%%%%%%%%%%%%%%%%%%%

The FIR-to-optical \oiii\ line ratio diagnostics presented in the left panel of Figure~\ref{fig:Fig3_OIII_Diagnostics} demonstrate that the observed values in \shortname\ cannot be explained by a homogeneous ionized gas with a single set of electron density and temperature. This discrepancy highlights the need to explore more complex scenarios for the ISM. In the following subsections, we discuss two possible origins of this tension: (i) the presence of inhomogeneously distributed dust within the ISM (\S~\ref{subsec:discussion_dust}) and (ii) a density-stratified ionized ISM structure (\S~\ref{subsec:discussion_two-phase} and \S~\ref{subsec:discussion_two-phase2}).

%%%%%%%%%%%%%%%%%%%%%%%%%%%%%%%%%%%%%%%%%%%%%%%%%%%%%%%%%%%%%%%
\subsection{Inhomogeneously Distributed Dust}\label{subsec:discussion_dust}
%%%%%%%%%%%%%%%%%%%%%%%%%%%%%%%%%%%%%%%%%%%%%%%%%%%%%%%%%%%%%%%

One possibility to explain the observed \oiii\ line ratios could be an inhomogeneous dust distribution. This idea is suggested by ISM studies of local dwarf galaxies using rest-frame optical to FIR emission lines (e.g., \citealp{Chen23, Kumari24, Chen24}). 

\cite{Chen24} analyzed both optical and FIR emission lines of \oiii\ (4364, 5008 \AA\ and 88 \micron) and \nii\ (5755, 6583 \AA\ and 122 \micron) in Haro\,3, to estimate both electron temperatures and chemical abundance ratios. 
They found a significant discrepancy between the N$^{+}$/H$^{+}$ and N$^{+}$/O$^{++}$ ratios derived from the optical and FIR \nii\ lines, which they interpreted as evidence for strong attenuation of optical lines due to inhomogeneously distributed dust.

In \cite{Chen24}, the dust extinction correction was applied using the Balmer decrement measured from spatially integrated fluxes. However, if dust is distributed inhomogeneously within the galaxy, such integrated measurements would be biased toward regions with less dust extinction, leading to an underestimation of the true dust extinction.

If this hypothesis holds for \shortname, the intrinsic \oiii\,5008 and 4364 fluxes would be underestimated. Equivalently, the intrinsic \oiii88/5008 and \oiii4364/5008 ratio would be overestimated, potentially affecting the derived ISM physical properties. We test this hypothesis for \shortname\ as follows.

First, we generate an $A_V$ map from the PSF-matched H$\alpha$/H$\beta$ ratio map within the $3\sigma$ regions of both line-intensity maps (right panel of Figure~\ref{fig:Fig3_OIII_Diagnostics}). We apply Voronoi pixel binning \citep{CappellariCopin03,Cappellari09} to the integrated intensity maps of H$\beta$ and H$\alpha$ in order to optimize the signal-to-noise ratio in each pixel. The observed \oiii4364 and \oiii5008 intensity maps are then corrected for dust extinction on a pixel-by-pixel basis using the resulting $A_V$ map\footnote{We note that the \oiii4364 detection region (cyan contours in Figure~\ref{fig:Fig3_OIII_Diagnostics}) is smaller than the $A_{\rm V}$ map coverage. The dust extinction correction is applied even to low-significance pixels in the \oiii4364 map. This could result in an underestimation of the dust-corrected \oiii4364 flux, and thus of the \oiii4364/5008 ratio. Nevertheless, even if the ratio is underestimated, the true value would shift upward along the vertical axis in Figure~\ref{fig:Fig3_OIII_Diagnostics}, which does not alter our conclusion.}.
Outside the region where the $A_{\rm V}$ map is not reliably defined (i.e., where either H$\alpha$ or H$\beta$ is not detected above $3\sigma$), we assume $A_{\rm V} = 0$ for simplicity.

We measure the total dust-corrected \oiii5008 and \oiii4364 fluxes in a similar manner to the observed total fluxes (Section~\ref{subsec:data1}). We perform isophotal photometry on both dust-corrected line intensity maps, where we use the $2\,\sigma$ isophotal aperture defined in the \oiii5008 map (the outermost gray contour in the right panel of Figure~\ref{fig:Fig3_OIII_Diagnostics}). Then, we applied a 10\% aperture correction (see \S\ \ref{subsec:data1}) to both line flux values. 

Finally, we estimate the pixel-by-pixel dust-corrected \oiii4364/5008 and \oiii88/5008 ratios, represented by a gray star in the left panel of Figure~\ref{fig:Fig3_OIII_Diagnostics}. As a result, we find that the \oiii4364/5008 and \oiii88/5008 ratios change by factors of $\sim 1.07$ and $\sim 0.82$, respectively. 
However, even after this correction, the integrated \oiii\ ratios of \shortname\ still cannot be explained by a homogeneous ionized gas with densities comparable to those inferred from the \oii\ lines. 

We therefore conclude that the integrated \oiii\ ratios cannot be reproduced by an ISM characterized by a single set of electron density and temperature, even after accounting for the spatial structure of dust attenuation. This supports the view that inhomogeneously distributed dust is not the primary cause of the observed discrepancy in the \oiii\ line ratios of \shortname.

%%%%%%%%%%%%%%%%%%%%%%%%%%%%%%%%%%%%%%%%%%%%%%%%%%%%%%%%%%%%%%%
\subsection{Evidence for a density-stratified ionized ISM}\label{subsec:discussion_two-phase}
%%%%%%%%%%%%%%%%%%%%%%%%%%%%%%%%%%%%%%%%%%%%%%%%%%%%%%%%%%%%%%%

%FFFFFFFFFFFFFFFFFFFFFFFFFFFFFFFFFFFFFFFFFFFFFFFFFFFFFFFFFFFFF%
\begin{figure*}[t]
    \begin{center}
        \includegraphics[width=0.8\linewidth, angle=0]{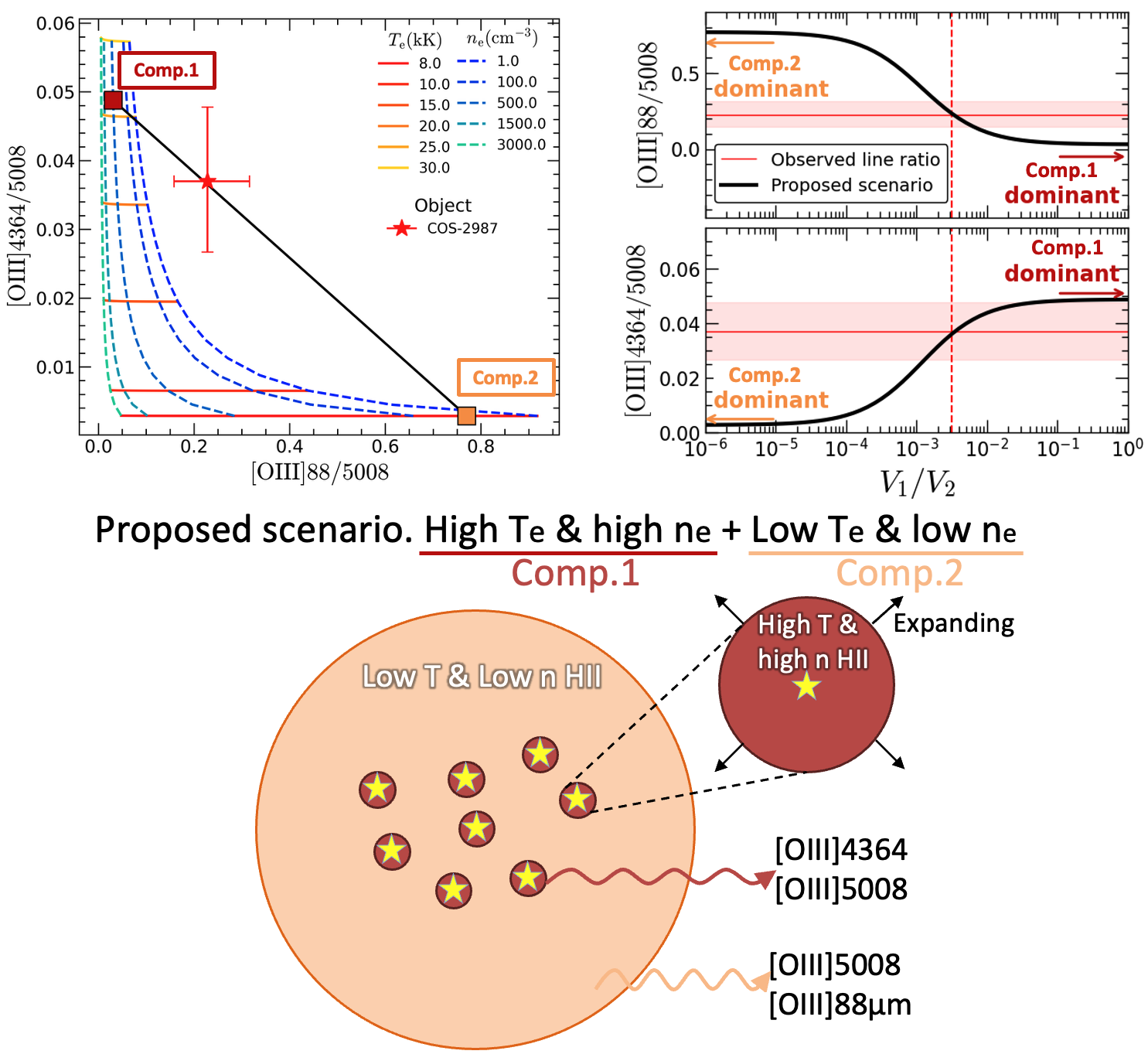}\\
    \end{center}
    \caption{
    (\textit{Top left}) Same as the left panel of Figure~\ref{fig:Fig3_OIII_Diagnostics}, but shown with linear scales on both axes. The two squares correspond to the two  gas components: the dark red square indicates component~1, with ($T_e$, $n_e$) $=$ (26,000~K, 500~cm$^{-3}$), and the orange square indicates component~2, with ($T_e$, $n_e$) $=$ (8,000~K, 50~cm$^{-3}$). The observed line ratios of \shortname\ (red star) can be explained by an ISM composed of these two components, mixed at an appropriate volume ratio and represented by the line connecting the two squares.
    (\textit{Top right}) 
    \oiii88/5008 and \oiii4364/5008 luminosity ratios as a function of volume ratio $V_{1} / V_{2}$ in our two-component model. The black curves show the predicted ratios from the model. The horizontal solid lines and shaded regions indicate the observed ratios and their $1\sigma$ uncertainties, respectively. The observed values are reproduced when $V_1 : V_2 \sim 1 : 300$ (vertical dashed lines).
    (\textit{Bottom}) Schematic illustration of our proposed two-component gas scenario. Hot and dense H\,{\sc ii} regions around young massive stars or clusters (component~1) are surrounded by relatively cool and diffuse H\,{\sc ii} regions (component~2). From the perspective of pressure balance, the component~1 gas is expected to be expanding. The \oiii4364 and \oiii88 emission arise primarily from the component~1 and 2 regions, respectively, while both components contribute to the \oiii5008 emission. 
    }
    \label{fig:Fig4_2PhaseModel}
\end{figure*}
%FFFFFFFFFFFFFFFFFFFFFFFFFFFFFFFFFFFFFFFFFFFFFFFFFFFFFFFFFFFFF%

A common assumption in deriving nebular physical parameters is that all emission lines originate from a homogeneous ionized gas with uniform electron density and temperature (see \citealt{Kewley19} and references therein). However, the critical density of \oiii88 ($n_{\rm e,crit} = 510~{\rm cm^{-3}}$) is approximately 3–4 orders of magnitude lower than those of \oiii4364 ($3.0 \times 10^6~{\rm cm^{-3}}$) and \oiii5008 ($6.8 \times 10^5~{\rm cm^{-3}}$) (e.g., \citealp{Osterbrock06, Draine11}). This large difference in critical densities suggests that the \oiii88, \oiii5008, and \oiii4364 lines may originate from physically distinct ionized gas conditions with different densities and temperatures.

We tested this possibility by examining the spatial and kinematic distributions of the \oiii5008 and \oiii88 lines.
As shown in Figure~\ref{apdx_fig:OIII_RadialProfile} (Appendix~\ref{appendix2}), the radial profiles of the two lines are consistent within uncertainties, indicating no significant spatial segregation at the current ALMA resolution. Similarly, Figure~\ref{apdx_fig:OIII_LineProfile} in Appendix~\ref{appendix3} shows that the spectral profiles of the two lines are broadly consistent, although the \oiii88 line appears marginally broader. These results do not provide evidence for spatial or kinematic separation between the emitting regions at the given spatial and spectral resolutions of the current data. Moreover, we do not find any velocity components that are detected in the \oiii88 line but absent in the \oiii5008 line, which would otherwise be expected if dust attenuation played a major role. This further supports our conclusion in \S~\ref{subsec:discussion_dust} that inhomogeneously distributed dust is not the primary cause of the observed discrepancy.

Motivated by the large differences in critical densities, we explore the possibility that the ionized gas in \shortname\ consists of multiple components with different electron densities and temperatures (i.e., a density-stratified ionized ISM). Although no significant spatial or kinematic offsets are observed, these findings do not rule out the presence of physically distinct gas. While comprehensive models for ionized gas with multiple physical conditions, such as \texttt{MULTIGRIS} \citep{Lebouteiller2022, Ramambason2022}, are available, we adopt a simplified two-component toy model that focuses on three emission lines from doubly ionized oxygen: \oiii4364, \oiii5008, and \oiii88. A similar analysis has been independently conducted for galaxies at $z = 6$ to 9 by \citet{Harikane25b}. By limiting the number of free parameters, our aim is to identify the key physical conditions responsible for the observed line ratios.

In a two-component ionized gas model, the luminosity of a given emission line ($L_{\rm line}$, where line refers to \oiii4364, \oiii5008 or \oiii88) can be expressed as:
{\small
\begin{align}
L_{\rm line} = \left( \frac{\epsilon_{\rm line,1}}{\rm erg~s^{-1}~cm^{-3}} \right) \left( \frac{V_1}{\rm cm^3} \right) + \left( \frac{\epsilon_{\rm line,2}}{\rm erg~s^{-1}~cm^{-3}} \right) \left( \frac{V_2}{\rm cm^3} \right),
\label{eq:two-phase_luminosity}
\end{align}
}
where $\epsilon_{{\rm line},i}$ and $V_i$ ($i=1$ or 2) are the volume emissivity and the volume of each ionized gas component, respectively. The emissivity $\epsilon_{{\rm line},i}$ is given by:
\begin{align}
\epsilon_{{\rm line},i} = \left( \frac{\Lambda_{{\rm line},i}(n_{{\rm e},i}, T_{{\rm e},i})}{\rm erg~s^{-1}~cm^{3}} \right) \left( \frac{n_{{\rm e},i}}{{\rm cm^{-3}}} \right) \left( \frac{n_{{\rm O^{++}},i}}{{\rm cm^{-3}}} \right),
\label{eq:volume_emissivity}
\end{align}
where $\Lambda_{{\rm line},i}$ is the cooling rate, and $n_{{\rm O^{++}},i}$ is the number density of doubly ionized oxygen. We calculate $\Lambda_{{\rm line},i}$ for each line using {\tt PyNeb} (\citealp{Luridiana15}) by assuming specific values of $n_{{\rm e},i}$ and $T_{{\rm e},i}$. We assume $n_{{\rm O^{++}},i} = n_{{\rm e},i} \times 10^{-4}$, corresponding to ${\rm O^{++}/H^{+}} = 10^{-4}$ (i.e., $12+\log({\rm O^{++}/H^{+}}) = 8$). The aim of this analysis is not to determine best-fit parameters, but rather to assess whether the observed \oiii\ luminosity ratios can be reproduced within the framework of a density-stratified ionized ISM.

In principle, the line luminosity ratios predicted by the two-component model can be expressed as a linear combination of the ratios corresponding to the individual components. Accordingly, in the top left panel of Figure \ref{fig:Fig4_2PhaseModel}, which is the same as Figure \ref{fig:Fig3_OIII_Diagnostics} but shown with linear scales on both axes, hot and dense gas (component~1) and relatively cool and diffuse gas (component~2) reside on the model grid representing a homogeneous ionized medium. The composite (i.e., global) line ratios are therefore expected to lie along the straight line connecting these two components.

We find that the observed \oiii\ line ratios are well reproduced by a model consisting of component~1 with $(T_{\rm e,1}, n_{\rm e,1}) = (26,000\,{\rm K}, 500\,{\rm cm^{-3}})$ and component~2 with $(T_{\rm e,2}, n_{\rm e,2}) = (8,000\,{\rm K}, 50\,{\rm cm^{-3}})$. The {components~1 and 2 are marked by dark red and orange squares in the top left panel of Figure~\ref{fig:Fig4_2PhaseModel}. 
As shown in the figure, the observed line ratios (indicated by the red star) can be reproduced by a linear combination of the two components with an appropriate volume ratio. We also note that other sets of physical conditions could reproduce the observed line ratios, such as a higher $T_{\rm e}$ for component~1 and a higher $n_{\rm e}$ for component~2.
The top right panel of Figure~\ref{fig:Fig4_2PhaseModel} shows the predicted line ratios of the two-component model as a function of the volume ratio $V_1/V_2$, with $T_{\rm e,1}$, $n_{\rm e,1}$, $T_{\rm e,2}$, and $n_{\rm e,2}$ held fixed.
By comparing the model predictions with the observed values of \oiii88/5008 and \oiii4364/5008, we derive a volume ratio of $V_1/V_2 \sim 1/300$. The corresponding mass ratio of the two gas components is $M_1/M_2 \sim 1/30$, where the mass of each component is calculated as $M_{\rm i} = n_{\rm e,i} V_{\rm i}$.

We stress again that the specific combination of $T_{\rm e}$, $n_{\rm e}$, and $V_1/V_2$ parameters presented above is not a unique solution.
Nevertheless, our two-component model successfully reproduce both the observed \oiii88/5008 and \oiii4364/5008 ratios simultaneously, that cannot be achieved by a commonly assumed homogeneous gas. We also note that the observed \oii3727,3730 emission likely originates from a different ionization phase not included in our current model. One possible explanation is the presence of a third component with lower ionization and relatively high density, perhaps located at the surface of H\,{\sc i} gas clouds in the galaxy. Exploring this additional phase is beyond the scope of the present work, but it will be an important direction for future modeling efforts.

%%%%%%%%%%%%%%%%%%%%%%%%%%%%%%%%%%%%%%%%%%%%%%%%%%%%%%%%%%%%%%%
\subsection{Interpretation of the the density-stratified ionized ISM}\label{subsec:discussion_two-phase2}
%%%%%%%%%%%%%%%%%%%%%%%%%%%%%%%%%%%%%%%%%%%%%%%%%%%%%%%%%%%%%%%

The bottom panel of Figure~\ref{fig:Fig4_2PhaseModel} provides a schematic illustration of the density-stratified ionized ISM scenario.
In this framework, component~1 corresponds to compact, dense, and high-temperature ionized regions associated with young massive stars or stellar clusters. Such regions are expected to be more common in high-$z$ galaxies \citep{Mingozzi22,Topping25}.
The relatively cool and diffuse gas (component~2) is expected to be surrounding the component~1 regions because their volumes are different by a factor of $\sim 300$. The diffuse and extended gas component may originate from
hydrogen-ionizing photons (i.e., LyC photons) leaking from dense, high-temperature {\sc H ii} regions (component~1), resembling the so-called ``picket-fence model'' \citep{Heckman11, Leitet13}. This scenario is supported by observations of nearby massive star-forming regions: {\it AKARI} observations of 30Doradus \citep{Kawada11} and {\it Herschel}/PACS observations of LMC-N11 \citep{Lebouteiller12} revealed spatially extended \oiii88\ emission beyond the regions expected from the number of massive stars, likely due to a clumpy, low-density medium that allows ionizing photons to propagate over large scales. Similarly, modeling of the {\it Herschel} Dwarf Galaxy Survey \citep{Cormier15, Cormier19, Ramambason22} also requires such a high-ionization, low-density component to reproduce the observed \oiii88 flux. Such porous conditions would facilitate the leakage of LyC photons into the surrounding medium, relevant to the study of cosmic reionization \citep[e.g.,][]{Cormier19, Ramambason22, Hagimoto25}.

Indeed, the detection of Ly$\alpha$ emission from \shortname\ at $z \sim 6.8$ \citep{Laporte17}, prior to the completion of cosmic reionization, supports the presence of clear sight lines that facilitate LyC escape. We refer readers to the estimate of the LyC photon escape fraction for \shortname\ presented by \cite{Mawatari25}. 
Using the analytic prescription of \cite{Choustikov24}, which incorporates parameters such as the O32 ratio (\oiii4960,5008/\oii3727,3730) and H$\beta$ luminosity, \citet{Mawatari25} obtained a non-zero escape fraction of $f_{\rm esc} = 5^{+8}_{-5}\%$. This estimate is consistent with the high \oiii88\micron-to-\cii158\micron\ luminosity ratio of $\approx 6$ in \shortname\ \citep{Witstok22}. Based on the empirical relation proposed by \cite{ura23}\footnote{\cite{ura23} combined two positive correlations: (i) between the optical O32 ratio and $f_{\rm esc}$, and (ii) between the optical O32 ratio and the \oiii88\micron-to-\cii158\micron\ luminosity ratio, thereby obtaining a correlation between the \oiii88\micron-to-\cii158\micron\ ratio and $f_{\rm esc}$.}, we infer $f_{\rm esc} \approx 4$--$14\%$. Taken together, the results for \shortname, along with similar trends in other high-$z$ galaxies (Figure~\ref{fig:Fig3_OIII_Diagnostics}), may hint that high-$z$ ALMA \oiii88\micron\ emitters—with potentially porous ionized gas structures—contributed to cosmic reionization.

To account for the extremely high electron temperature ($T_{\rm e} \sim 20,000$ K) and the resulting strong \oiii4364 emission of component~1, we note that one speculative possibility is the presence of an extreme stellar population, such as very massive stars \citep[VMSs with $M \gtrsim 150,M_\odot$;][]{Crowther10,Vink15}. Future follow-up observations in the rest-frame UV may provide useful clues by examining spectral features that could be indicative of VMSs, such as an enhanced N/O ratio and a strong He~{\sc ii}\,1640 equivalent width \citep{Vink23,Upadhyaya24,Senchyna24}.

%%%%%%%%%%%%%%%%%%%%%%%%%%%%%%%%%%%%%%%%%%%%%%%%%%%%%%%%%%%%%%%
\section{Conclusions}\label{sec:conclusion}
%%%%%%%%%%%%%%%%%%%%%%%%%%%%%%%%%%%%%%%%%%%%%%%%%%%%%%%%%%%%%%%

In this Letter, we investigated the ionized interstellar medium (ISM) of the $z=6.8$ galaxy \shortname\ using JWST/NIRSpec IFS and ALMA spectroscopy. Our main conclusions are summarized as follows:
\begin{itemize}
    \item[(1)] 
    JWST/NIRSpec spectra revealed multiple rest-frame optical emission lines in \shortname, including \oii~$\lambda\lambda$~3727, 3730, \oiii~4364, \oiii~$\lambda\lambda$~4960, 5008, \oiii\ 88 \micron, as well as H$\alpha$ and H$\beta$ (Figure~\ref{fig:fig1_OIII_3lines}; Table~\ref{tab:SpatiallyIntegratedProp}).  
    Using only the JWST/NIRSpec lines, we derived an electron density of $n_{\rm e} \sim 500$ cm$^{-3}$ from the \oii\ doublet, consistent with the redshift evolution trend of $n_{\rm e}$ reported in previous studies (Figure~\ref{fig:Fig2_z_vs_ne}). 
    \item[(2)] 
    From the combination of optical and FIR \oiii\ lines, we constructed a diagnostic diagram based on the \oiii88/\oiii5008 and \oiii4364/\oiii5008 ratios. COS-2987 falls well outside the grid predicted by homogeneous models computed with {\tt PyNeb}.
    Furthermore, a compilation of other high-$z$ galaxies shows a similar behavior, in which the $n_{\rm e}$ values inferred from the two combinations of \oiii, \oiii88/\oiii5008 and \oiii4364/\oiii5008, are systematically lower than those derived from optical diagnostics such as \oii. This indicates that the discrepancy is likely a general trend, and demonstrates that the assumption of a uniform ionized gas with a single set of electron density and temperature could not be valid in a large fraction of high-$z$ galaxies with strong \oiii88 (Figure~\ref{fig:Fig3_OIII_Diagnostics}).
    \item[(3)] 
    We examined whether inhomogeneously distributed dust could explain the high \oiii88/\oiii5008 ratios. 
    We conclude that a clumpy dust distribution is not the primary cause of the discrepancy (see \S~\ref{subsec:discussion_dust} and the right panel of Figure~\ref{fig:Fig3_OIII_Diagnostics}).  
    \item[(4)] 
    Instead, we showed that the observations can be reproduced by invoking a density-stratified ionized ISM. A simple two-component ionized-gas toy model with $(T_{\rm e}, n_{\rm e}) \sim (26{,}000\,{\rm K}, 500\,{\rm cm^{-3}})$ and $(8{,}000\,{\rm K}, 50\,{\rm cm^{-3}})$ successfully reproduces the observed \oiii\ ratios.
    We discussed a physical scenario for such stratification: porous ISM conditions in which LyC photons leak from dense regions into a low-density medium, akin to a ``picket-fence'' geometry (\S~\ref{subsec:discussion_two-phase} and \S~\ref{subsec:discussion_two-phase2}; Figure~\ref{fig:Fig4_2PhaseModel}).
\end{itemize}

In summary, these results highlight the complex ionized gas structure of the star-forming galaxy \shortname, which may reflect its clumpy and extended morphology \citep{Mawatari25}. Given that such clumpy, multi-component stellar structures are not uncommon among high-redshift galaxies (e.g., \citealt{Sugahara25, Harikane25a}), a future systematic investigation into the ubiquity of density-stratified ionized ISM at high redshift would be of great interest.  
A more detailed modeling of the multi-component ISM in a larger sample of high-$z$ galaxies will be addressed in a forthcoming paper (Y. Sugahara et al., in preparation).  
Given their high critical densities, rest-frame UV emission lines such as O~\textsc{iii}] and C~\textsc{iii}] offer a promising avenue for probing density stratification in high-$z$ galaxies. In addition, high-angular-resolution observations with ALMA are essential for obtaining spatially resolved measurements of electron temperature, electron density, and the \oiii88/5008 ratio, thereby enabling constraints on the structure and physical conditions of density-stratified ionized ISM.

%%%%%%%%%%%%%%%%%%%%%%%%%%%%%%%%%%%%%%%%%%%%%%%%%%%%%%%%%%%%%%%
\begin{acknowledgments}
%%%%%%%%%%%%%%%%%%%%%%%%%%%%%%%%%%%%%%%%%%%%%%%%%%%%%%%%%%%%%%%

e are grateful to an anonymous referee for valuable comments that have greatly improved the paper.
We thank Masami Ouchi, Yoshiaki Ono, and Kimihiko Nakajima for providing us with valuable comments on the presentations of the first author.
We also acknowledge Joris Witstok, Fengwu Sun, Tohru Nagao, Hide Yajima, Hajime Fukushima, Nario Kuno, Shunsuke Honda, Asahi Hamada, Yuuki Takagishi, and Yuzuru Terui for insightful discussion.
This paper makes use of the following ALMA data: ADS/JAO.ALMA\#2018.1.00429.S. ALMA is a partnership of ESO (representing its member states), NSF (USA), and NINS (Japan), together with NRC (Canada), MOST and ASIAA (Taiwan), and KASI (Republic of Korea), in cooperation with the Republic of Chile. The Joint ALMA Observatory is operated by ESO, AUI/NRAO, and NAOJ.
This work has made use of data from the European Space Agency (ESA) mission {\it Gaia} (\url{https://www.cosmos.esa.int/gaia}), processed by the {\it Gaia} Data Processing and Analysis Consortium (DPAC, \url{https://www.cosmos.esa.int/web/gaia/dpac/consortium}).Funding for the DPAC has been provided by national institutions, in particular the institutions participating in the {\it Gaia} Multilateral Agreement. The {\it Gaia} data are retrieved from the JVO portal (\url{http://jvo.nao.ac.jp/portal}) operated by the NAOJ. 
This research has made use of NASA’s Astrophysics Data System.
T.H. was supported by Leading Initiative for Excellent Young Researchers, MEXT, Japan (HJH02007) and by JSPS KAKENHI grant Nos. 22H01258, 23K22529, and 25K00020. K.M. acknowledges financial support from JSPS through KAKENHI grant No. 20K14516. K.M. and A.K.I are supported by JSPS KAKENHI grant No. 23H00131. J.A.-M., L.C. and S.A. acknowledge support by grants PIB2021-127718NB-100 from the Spanish Ministry of Science and Innovation/State Agency of Research MCIN/AEI/10.13039/501100011033 and by “ERDF A way of making Europe”. 
J.A.-M. and C. B-P acknowledge support by grant PID2024-158856NA-I00 from the Spanish Ministry of Science and Innovation/State Agency of Research MCIN/AEI/10.13039/501100011033 and by “ERDF A way of making Europe”.
Y.N. acknowledges funding from JSPS KAKENHI Grant Number 23KJ0728. A.C.G. acknowledges support by JWST contract B0215/JWST-GO-02926. Y.W.R. was supported by JSPS KAKENHI Grant Number 23KJ2052. Y.F. acknowledges supports from JSPS KAKENHI Grant Numbers JP22K21349 and JP23K13149. M.H. was supported by JSPS KAKENHI Grant No. 22KJ1598. 
The project that gave rise to these results received the support of a fellowship from the “la Caixa” Foundation (ID 100010434). The fellowship code is LCF/BQ/PR24/12050015. L.C. acknowledges support from grants PID2022-139567NB-I00 and PIB2021-127718NB-I00 funded by the Spanish Ministry of Science and Innovation/State Agency of Research  MCIN/AEI/10.13039/501100011033 and by “ERDF A way of making Europe”. D.C. is supported by research grant PID2021-122603NB-C21 funded by the Ministerio de Ciencia, Innovación y Universidades (MI-CIU/FEDER) and the research grant CNS2024-154550 funded by MI-CIU/AEI/10.13039/501100011033. C.B.P. acknowledges the support of the Consejería de Educación, Ciencia y Universidades de la Comunidad de Madrid through grants No. PEJ-2021-AI/TIC-21517 and PIPF-2023/TEC29505. 
\end{acknowledgments}
%%%%% Software %%%%%
\software{scipy, astropy, PyNeb, photutils}
\clearpage

%%%%%%%%%%%%%%%%%%%%%%%%%%%%%%%%%%%%%%%%%%%%%%%%%%%%%%%%%%%%%%%
\appendix
%%%%%%%%%%%%%%%%%%%%%%%%%%%%%%%%%%%%%%%%%%%%%%%%%%%%%%%%%%%%%%%

%%%%%%%%%%%%%%%%%%%%%%%%%%%%%%%%%%%%%%%%%%%%%%%%%%%%%%%%%%%%%%%
\section{Point Spread Function of the NIRSpec IFU Data Cube}\label{appendix1}
\setcounter{figure}{0}% Reset figure counter
\renewcommand{\thefigure}{A.\arabic{figure}} % Figures numbered as A.1, A.2, etc.
%%%%%%%%%%%%%%%%%%%%%%%%%%%%%%%%%%%%%%%%%%%%%%%%%%%%%%%%%%%%%%%

To ensure that the line fluxes are measured from the same spatial region, we used a point spread function (PSF)-matched data cube, as described in \cite{Mawatari25}. Here, we briefly summarize the procedure used to create the PSF-matched data cube.
To obtain an empirical PSF for the data cube, we analyzed observations of the A3V star 1808347 (2MASS J18083474+6927286) taken during the commissioning program (PID 1128; PI: N. L\"{u}tzgendorf; \citealp{Boker23}), using the same grating/filter combination (G395H/F290LP) as for \shortname. The data reduction for the standard star was performed with the same JWST pipeline version and CRDS context as for \shortname.

From the data cube of the A3V star, we generated PSF images at the wavelengths corresponding to the emission lines detected in \shortname. We then created convolution kernels to homogenize all PSFs to match the PSF at $5.125\,\mu$m, corresponding to the observed wavelength of H$\alpha$ in \shortname, which has the largest FWHM of $0\farcs21$ among the detected lines.

%%%%%%%%%%%%%%%%%%%%%%%%%%%%%%%%%%%%%%%%%%%%%%%%%%%%%%%%%%%%%%%
\section{Radial Profiles of \oiii5008 and \oiii88}\label{appendix2}
\setcounter{figure}{0}% Reset figure counter
\renewcommand{\thefigure}{B.\arabic{figure}} % Figures numbered as B.1, B.2, etc.
%%%%%%%%%%%%%%%%%%%%%%%%%%%%%%%%%%%%%%%%%%%%%%%%%%%%%%%%%%%%%%%

%FFFFFFFFFFFFFFFFFFFFFFFFFFFFFFFFFFFFFFFFFFFFFFFFFFFFFFFFFFFFF%
\begin{figure}[b]
    \begin{center}
        \includegraphics[width=14cm]{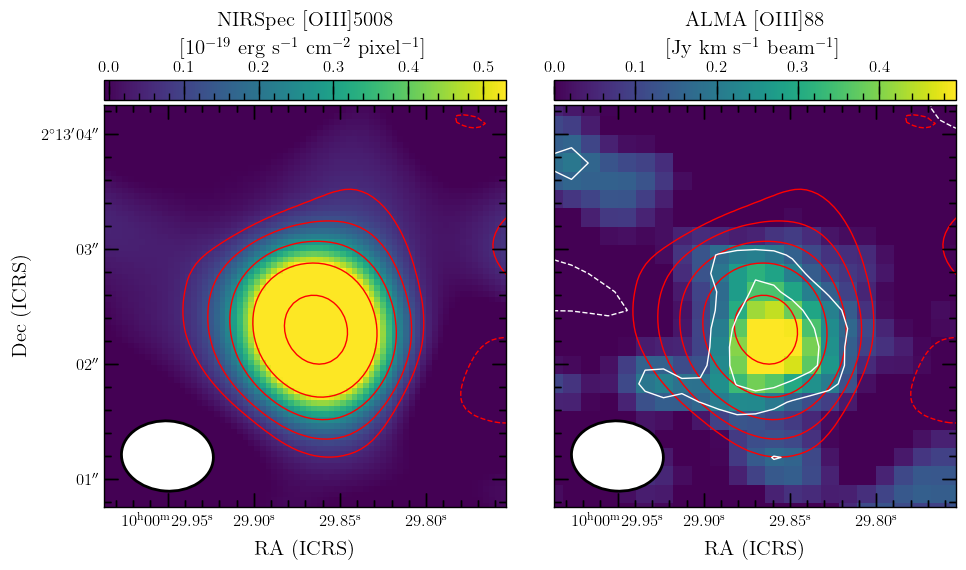}\\
    \end{center}
    \vspace{-0.5cm}
    \caption{
    (\textit{Left}) Integrated intensity map of the NIRSpec \oiii5008 emission, after convolution to match the ALMA beam size ($0\farcs80 \times 0\farcs61$). Red contours show the $\pm2^{n}\sigma$ significance levels ($n = 1, 2, 3, \dots$), where $\sigma = 1.16 \times 10^{-20}\,{\rm erg\,s^{-1}\,cm^{-2}\,pixel^{-1}}$. The orange circle at the bottom left indicates the FWHM of $0\farcs21$ at the observed wavelength of H$\alpha$ ($5.125\,\mu$m).
    (\textit{Right}) Integrated intensity map of the ALMA \oiii88 emission. White contours show the $\pm(2, 4, 6)\sigma$ significance levels, where $\sigma = 82.6\ {\rm mJy\,beam^{-1}km\,s^{-1}}$. Positive and negative contours are shown by the white solid and dashed lines, respectively.
    The white ellipse at the bottom left represents the synthesized beam size of $0\farcs80 \times 0\farcs61$.
    }
    \label{apdx_fig:OIII_SpatialDistribution}
\end{figure}
%FFFFFFFFFFFFFFFFFFFFFFFFFFFFFFFFFFFFFFFFFFFFFFFFFFFFFFFFFFFFF%

We examined the spatial distributions of the \oiii5008 and \oiii88 emission lines, as shown in Figures~\ref{apdx_fig:OIII_SpatialDistribution} and \ref{apdx_fig:OIII_RadialProfile}. In Figure~\ref{apdx_fig:OIII_SpatialDistribution}, we present the integrated intensity maps of both lines, where the \oiii5008 map has been convolved to match the ALMA beam size ($0\farcs8 \times 0\farcs6$).
To quantitatively assess the spatial difference, we compare the radial profiles of \oiii5008 and \oiii88 in Figure~\ref{apdx_fig:OIII_RadialProfile}. The blue and red lines with shaded regions represent the radial profiles of \oiii5008 and \oiii88, respectively. The two emission lines exhibit consistent radial distributions within the uncertainties, indicating that there is no significant evidence of spatial separation between the two lines at the current ALMA resolution.

We note, however, that this result does not exclude the possibility that future high-angular-resolution observations could reveal subtle morphological differences between the \oiii5008 and \oiii88 emissions.

%FFFFFFFFFFFFFFFFFFFFFFFFFFFFFFFFFFFFFFFFFFFFFFFFFFFFFFFFFFFFF%
\begin{figure}[t]
    \begin{center}
        \includegraphics[width=10cm]{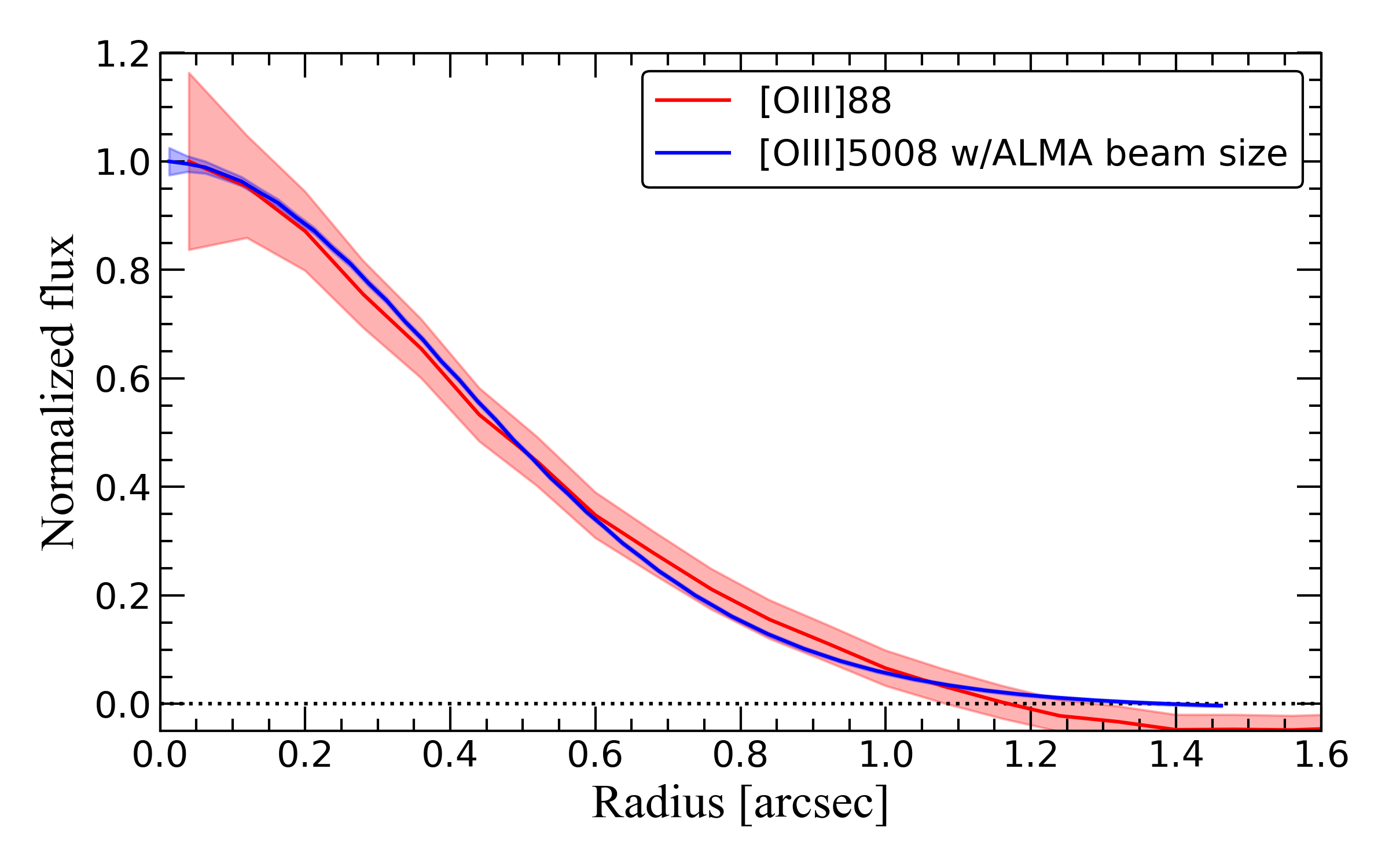}\\
    \end{center}
    \vspace{-0.5cm}
    \caption{
    Radial profiles of the \oiii5008 and \oiii88 emission lines. The blue and red lines represent the profiles of \oiii5008 and \oiii88, respectively, with shaded regions indicating their corresponding uncertainties.
    }
    \label{apdx_fig:OIII_RadialProfile}
\end{figure}
%FFFFFFFFFFFFFFFFFFFFFFFFFFFFFFFFFFFFFFFFFFFFFFFFFFFFFFFFFFFFF%

%%%%%%%%%%%%%%%%%%%%%%%%%%%%%%%%%%%%%%%%%%%%%%%%%%%%%%%%%%%%%%%
\section{Line Profiles of \oiii5008 and \oiii88}\label{appendix3}
\setcounter{figure}{0} % Reset figure counter
\renewcommand{\thefigure}{C.\arabic{figure}} % Figures numbered as C.1, C.2, etc.
%%%%%%%%%%%%%%%%%%%%%%%%%%%%%%%%%%%%%%%%%%%%%%%%%%%%%%%%%%%%%%%

%FFFFFFFFFFFFFFFFFFFFFFFFFFFFFFFFFFFFFFFFFFFFFFFFFFFFFFFFFFFFF%
\begin{figure}[h]
    \begin{center}
        \includegraphics[width=12cm]{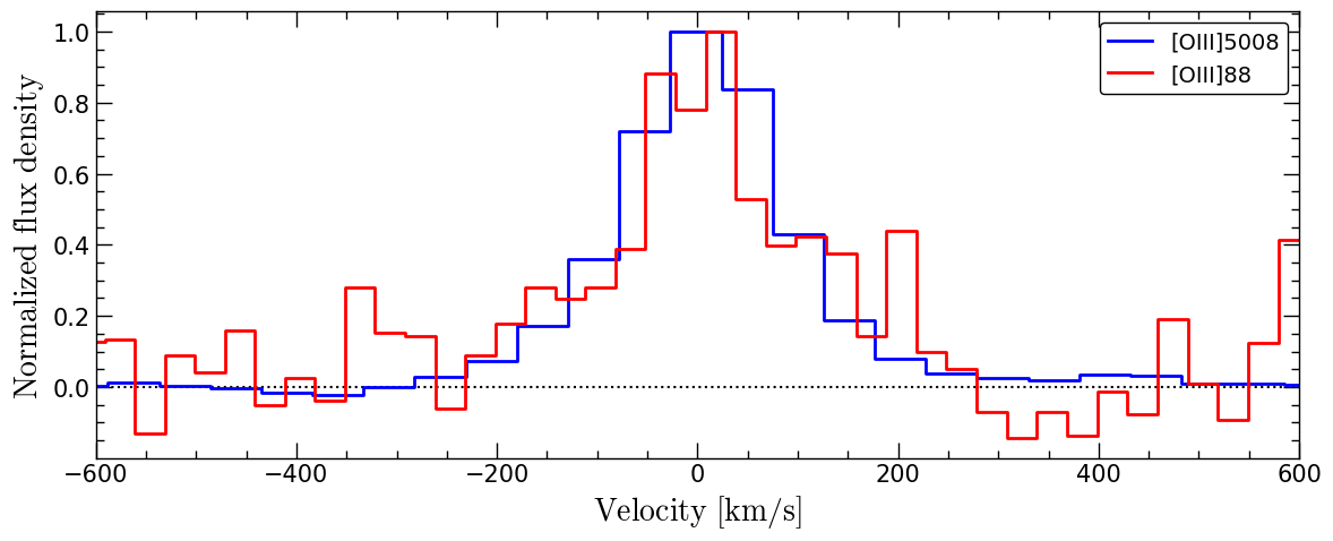}\\
    \end{center}
    \vspace{-0.5cm}
    \caption{
    Comparison of the velocity profiles of the \oiii5008 and \oiii88 emission lines. The blue and red lines represent the profiles of \oiii5008 and \oiii88, respectively, normalized to their peak intensities.
    }
    \label{apdx_fig:OIII_LineProfile}
\end{figure}
%FFFFFFFFFFFFFFFFFFFFFFFFFFFFFFFFFFFFFFFFFFFFFFFFFFFFFFFFFFFFF%

We investigated potential differences in the line profiles between the \oiii5008 and \oiii88 emissions. Figure~\ref{apdx_fig:OIII_LineProfile} shows the velocity profiles of \oiii5008 and \oiii88, represented by the blue and red lines, respectively.
Fitting each line with a single Gaussian profile and correcting for instrumental broadening, we obtain FWHM$_{5008} = 146 \pm 2$\,km\,s$^{-1}$ and FWHM$_{88} = 239 \pm 45$\,km\,s$^{-1}$. These values are consistent within $\sim2\sigma$, although the \oiii88 line appears marginally broader.

%%%%%%%%%%%%%%%%%%%%%%%%%%%%%%%%%%%%%%%%%%%%%%%%%%%%%%%%%%%%%%%

\bibliographystyle{aasjournal} 
%\bibliography{Usui24}

\end{document}